\documentclass[%
 reprint,
 amsmath,amssymb,
 aps,
]{revtex4-2}
\usepackage{graphicx}
\usepackage{dcolumn}
\usepackage{bm}
\usepackage{xcolor}
\usepackage{hyperref}
\usepackage{graphicx}
\usepackage[normalem]{ulem}
\usepackage{float}

\preprint{APS/123-QED}

\begin{document}
\title{Solitary death in coupled limit-cycle oscillators with higher-order interactions}
\author{Subhasanket Dutta}
\author{Umesh Kumar Verma}
\author{Sarika Jalan}\email{sarika@iiti.ac.in}
\affiliation{Complex Systems Lab, Department of Physics, Indian Institute of Technology Indore, Khandwa Road, Simrol, Indore-453552, India}

\date{\today}

\begin{abstract}
Coupled limit cycle oscillators with pairwise interactions depict phase transitions to amplitude or oscillation death. This Letter introduces a scheme for higher-order interactions, which can not be decomposed into pairwise interactions. We investigate Stuart Landau oscillators' dynamical evolution under the impression of such a coupling scheme and discover a particular type of oscillator death where a coupling-dependent stable death state, away from the origin, arises in isolation without being accompanied by any other stable state.  We call such a state a Solitary death state.  Moreover, the explosive transition to the death state is preceded by a surge in amplitude, followed by the revival of the oscillations. Such versatile dynamical states are further enriched with sensitivity to initial conditions. Finally, we point out the resemblance of the results with different dynamical states associated with epileptic seizures.

\end{abstract}


\maketitle
 \paragraph{{\bf Introduction:}}  Suppression of oscillations in dynamical systems has been an area of persistent interest due to its occurrence  in a wide range of real-world dynamical systems such as climate  \cite{climate}, Laser  \cite{laser}, electronic circuits 
 \cite{banerjee_exp}, cell differentiation 
 \cite{cell}, etc. Quenching of oscillations in large-scale dynamical systems made of interacting units arises primarily from  the coupling between these units.
 For instance, in Lasers, a few specific forms of the couplings among the laser components can lead to the quenching of oscillation \cite{laser}. In neurological systems, oscillation death has been proposed to be an important root cause of various neurodegenerative diseases and has been modeled using coupled nonlinear oscillators \cite{sanket, prasad_review}. 
 Coupled Stuart-Landau (SL) oscillators provide a prototype model to fathom the origin of oscillation death and associated changes in the stability properties. Earlier investigations on coupled SL oscillators trace to varieties of reasons behind the oscillation quenching, such as time delay \cite{koseska_review}, conjugate coupling \cite{hens_pal_dana}, and frequency mismatch \cite{koseska}. A death state of an oscillator can be classified into two major categories, amplitude death (AD) and oscillation death (OD) based on the spatial position and symmetry of the associated fixed points stability properties. AD state corresponds to all the oscillators settling down to the same fixed point, the unstable fixed point of the uncoupled oscillator. A coupled system stabilises the AD through Hopf bifurcation while preserving the symmetry \cite{koseska_review,prasad_review}. In contrast, in the OD state oscillators settle at different fixed points which originate due to coupling and a symmetry-breaking bifurcation \cite{banerjee_pre}. Further, there could be two different routes from the oscillatory state to the oscillation death state, a smooth second-order transition \cite{strogatz_ad,koseska,sanket} or an abrupt first-order jump \cite{verma1,laxmanan}.
 
 Furthermore, recently it has been increasingly realized that real-world complex systems made of dynamical units may not only have pairwise interactions but also possess higher-order structures; examples include cliques in the human brain \cite{brain_eg}, scientific collaborations \cite{multilayer_author} etc. 
 Studies of coupled Kuramoto oscillators with higher-order interactions have revealed various emerging behaviours, such as infinite multi-stable synchronized states and phenomena like abrupt (de)synchronization \cite{tanaka, skardal,JalanPRE2022}.
 The Kuramoto oscillator is a phase oscillator, whereas many real-world complex systems must be described with both amplitude and phase. SL oscillator is a limit cycle oscillator which takes this factor into account. 
 
 Some of the first results on pairwise coupled Stuart Landau oscillators on large networks was the manifestation of amplitude death at large coupling strength \cite{strogatz_ad, bard, yamaguchi}, with a spread in the intrinsic frequency of the oscillators causing a damping effect yielding amplitude death.

 \begin{figure*}
\includegraphics[height=8.0cm, width=0.75\linewidth]{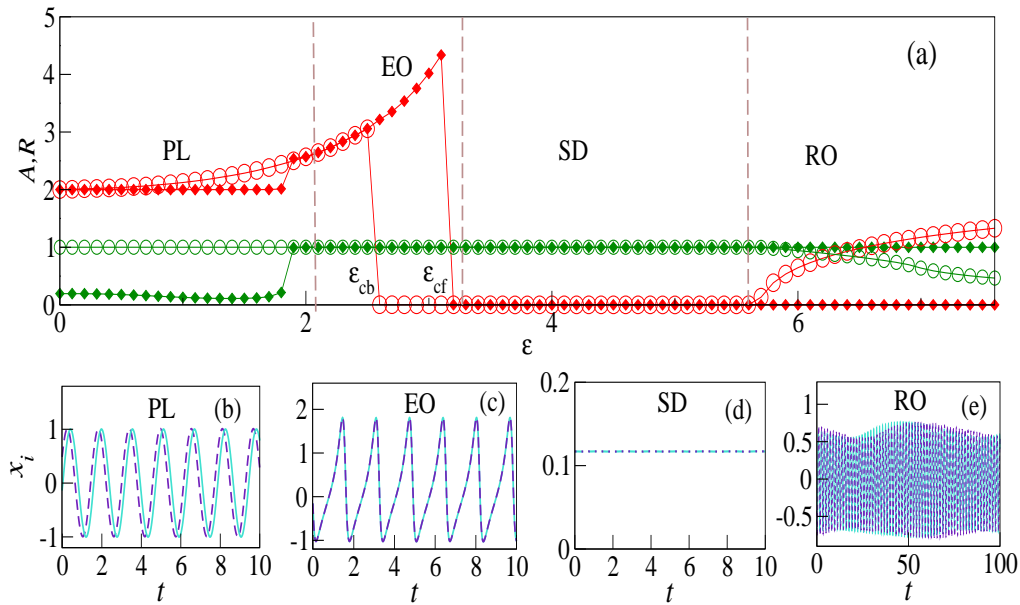}
    \caption{(a) $A(R)$ vs $\varepsilon$, (b)-(e)
 time-series for a system of identical globally coupled oscillators  (Eq.~\ref{higher_order_sl}), (b) Phase-locked (PL) ($\varepsilon=1.0$), (c) synchronized state with enhanced oscillation (EO) ($\varepsilon=2.3$), (d) solitary death (SD)   ($\varepsilon=3.4$), (e) revival of oscillation (RO) state with toroid ($\varepsilon=7.5$). Red diamond - $A$ in the forward direction, red circle - $A$ in the backward direction, green diamond -$R$ in the forward direction, green circle- $R$ in the backward direction. Other parameters are  $N=1000$, $\omega=4.0$, $\delta=1.1$.}
    \label{fig:figure1}
\end{figure*}
 Later, for identical oscillators, it was reported that time delay could result in the manifestation of AD \cite{delay_prl}.  
  Among other sources of AD for identical oscillators, Kenji \textit{et. al.} showed that dynamical coupling could also lead to AD, and Prasad \textit{et. al.} observed that when a system of Hindmarsh-Rose neurons oscillators were interacting via non-linear coupling, for sufficiently strong coupling strength death state can be reached \cite{kenji, prasad_nonlinear}. In these systems, dynamical and non-linear coupling, respectively, played the role of damping. Further,  the OD state was achieved by coupling two oscillators via only the real part \cite{koseska}.  Other types of couplings, such as conjugate \cite{chaos_2021, hens_pal_dana} and dissimilar \cite{sanket,prasad_diss} also resulted in the OD state along with the AD. Oscillator death is desired in many real-world systems having unwanted oscillations. For example, instability in the signals of Laser systems can be regulated via the amplitude death mechanism \cite{laser}. 
  In addition to the quenching, SL oscillators with pairwise couplings have been shown to depict a variety of rich behaviours, like synchronization \cite{sl_sync}, Chimera and Chimera death \cite{chimera_death}. 
  
 Recently, Carletti \textit{et al.} investigated coupled SL oscillators with linear higher-order interactions on networks \cite{carletti}. Note that the form of higher-order interactions considered in Ref.~\cite{carletti} decomposes into pairwise interactions without a network structure.
This Letter considers coupled Stuart Landau oscillators with higher-order non-linear multiplicative coupling which can not be decomposed into pairwise interactions. 
We find a first-order transition to synchronization, oscillator death and revival of the oscillations after the death state. 
A surge in the amplitude of the dynamical variable accompanies the abrupt transition to the synchronization state from the phase-locked state. Importantly, the oscillator death observed here does not manifest in the pairwise coupled SL cases.
An amplitude death is a symmetry state that arises when an unstable fixed point of the uncoupled system becomes stable due to the coupling.
Here, we report another origin of the symmetry-preserving state, which is the birth of a pair of stable-unstable fixed points through the saddle-node bifurcation in  SL oscillators coupled with triadic interactions. This pair of fixed points did not exist before the critical coupling strength, and its birth does not change in the stability properties of the already existing unstable fixed  point of the system. 
We refer to such a death state as Solitary death (SD) state to distinguish it from other coupling-created states which occur in more than one in number and are usually considered AD. Moreover, the property which separates SD from AD is the presence of bi-stability with a stable limit cycle which generally occurs with the OD state.
In this Letter, we perform linear stability analysis to find the criteria for the occurrence of the SD state. Also, we analyze the basin of attraction of the bi-stable regions during the first-order transition to synchronization and death states and draw bifurcation plots for our system. Finally, we check the robustness of the occurrence of all the phenomena against change in the value of the intrinsic frequency by introducing noise in  the initial conditions.

\paragraph{\bf{Model:}}
The dynamical equation for an uncoupled SL oscillator can be written as,
\begin{equation}
\label{uncoupled_sl}
	\dot z(t)=(a^2-|z(t)|^2)z+i\omega z
\end{equation}
Here $z$ is a complex variable depicting the dynamical state of an oscillator with $\omega$ being its intrinsic frequency. The oscillator has one unstable fixed point acting as a centre for a stable circular limit cycle of radius $a$.
We propose a coupling scheme for incorporating higher-order interactions among dynamical units. Our prime consideration while proposing the scheme is  that it should not be decomposed into pairwise terms. One of the simplest ways of satisfying this condition is to consider the product of the dynamical states of the interacting oscillators. Moreover, we avoided the conjugate variable ($z^*$) in the coupling function since it already yields quenching of the oscillations for pairwise coupling \cite{laxmanan}. Hence, it will be difficult to assess if the particular types of oscillations' quenching state reported in this Letter arises due to higher-order or conjugate couplings. However, the feedback coupling through $z_k$ in pairwise interaction does not result in quenching. 
Further, when transformed to polar coordinates, Eq.~1 signifies periodic coupling between the phases of the interacting oscillator, just like the form of higher-order coupling used in lower dimensional counterpart (Kuramoto oscillator) \cite{skardal} of SL oscillators.
\begin{equation}
\label{higher_order_sl}
	\dot z_j(t)=(1-|z_j(t)|^2)z_j+i\omega z_j + \frac{\varepsilon}{N^2}\sum_{k=1}^{N} \sum_{l=1}^{N} z_k z_l
\end{equation}
 \begin{figure}
\includegraphics[height=7.0cm,width=0.95\linewidth]{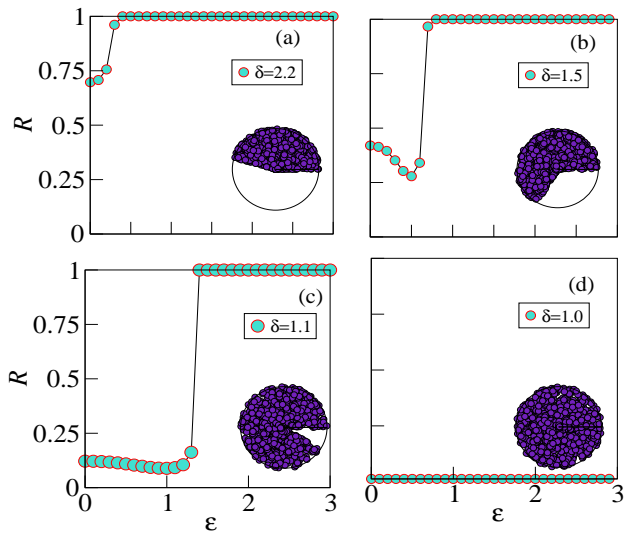}
    \caption{$R$ vs $\varepsilon$ for different initial conditions for a system of globally coupled identical SL oscillator (Eq.~\ref{higher_order_sl}) for $N=1000$. (a) $\delta=2.2$ (b) $\delta=1.5$, (c) $\delta=1.1$ (d) $\delta=1.0$. Here, the subplot represents the distribution of phase $\theta_i$ in the phase space ($r,\theta$).}
    \label{fig:figure2}
\end{figure}
Upon substituting $z_j=r_j e^{i\theta_j}$,  we get,
 \begin{eqnarray}
     \dot r_j &=&(1-r_j^2)r_j+\frac{\varepsilon}{N^2}\sum_{k,l=1}^{N}r_k r_l \cos(\theta_k+\theta_l-\theta_j)\nonumber \\
      \dot \theta_j &=&\omega_j+\frac{\varepsilon}{N^2 r_j}\sum_{k,l=1}^{N}r_k r_l \sin(\theta_k+\theta_l-\theta_j)
 \end{eqnarray}
where $r$ and $\theta$ are the amplitude and phase of the oscillator, respectively.
Upon substituting $z_j=x_j+iy_j$, the resulting equation is,
\begin{eqnarray}
\dot x_j&=& P^x_j+\frac{\varepsilon}{N^2}  \sum_{k,l=1}^{N}(x_k x_l-y_l y_k),\nonumber\\
 \dot y_j&=&P^y_j +\frac{\varepsilon}{N^2}  \sum_{k,l=1}^{N}(x_k y_l+x_l y_k)
 \label{eqn:SL_high}
\end{eqnarray}
where,
\begin{equation}
P^x_j=(1-x_j^2-y_j^2)x_j-w y_j,\: \: P^y_j=(1-x_j^2-y_j^2)y_j+w x_j
\nonumber
\end{equation}
We further define an order parameter ($A$) that quantifies the variance of fluctuation of the dynamical variables over a time span and tends to $0$ for the amplitude death. Moreover, to understand phase coherence, we use another order parameter, $R$, which takes 1 for the synchronized state and 0 for the incoherent state.
\begin{equation} \label{eqn:order1}
    A= \frac{1}{N}\sum_{i=1}^{N} (\langle x_i \rangle _{max,t}- \langle x_i \rangle _{min,t}),\
    R=\Bigg|\frac{\sum_{i=1}^{N} e^{i\theta_i}}{N}\Bigg|
    \nonumber
\end{equation}


\begin{figure}
\includegraphics[width=0.8\columnwidth]{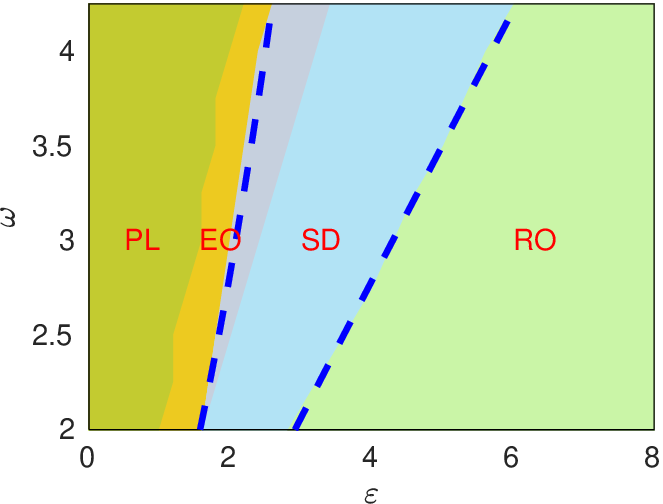}
    \caption{Phase diagram in the parameter space ($\varepsilon,\omega$) for a globally coupled identical SL oscillators system. The blue dashed line is obtained from analytical calculations that match the numerical results. The other parameters are $\delta=1.1$ and $N=1000$.}
    \label{fig:phase_space}
\end{figure}
\paragraph{\bf{Different dynamical states:}}
 The population of SL oscillators system coupled via higher-order interactions is affluent in dynamics and manifests several distinct dynamical states. The model considered here displays four distinct states namely phase-locked (PL), enhancement of oscillation (EO), solitary death (SD), and revival of oscillation (RO) states.  These states may arise in isolation or co-exist creating bi-stable regions in the phase space and having their own basin of attractions. 
 In the numerical simulations, for the forward direction, we consider uniform random initial conditions ($r \in[0,1]$ and $\theta \in [0, 2\pi/\delta]$) and increase the coupling strength adiabatically.  While in the backward direction, we start with random initial conditions and decrease the coupling value adiabatically. This was deliberately done to achieve the revival state, which guides us to how the fixed point losses its stability. We discuss all the states one by one.

\begin{figure}
\includegraphics[height=8.0cm,width=0.99\linewidth]{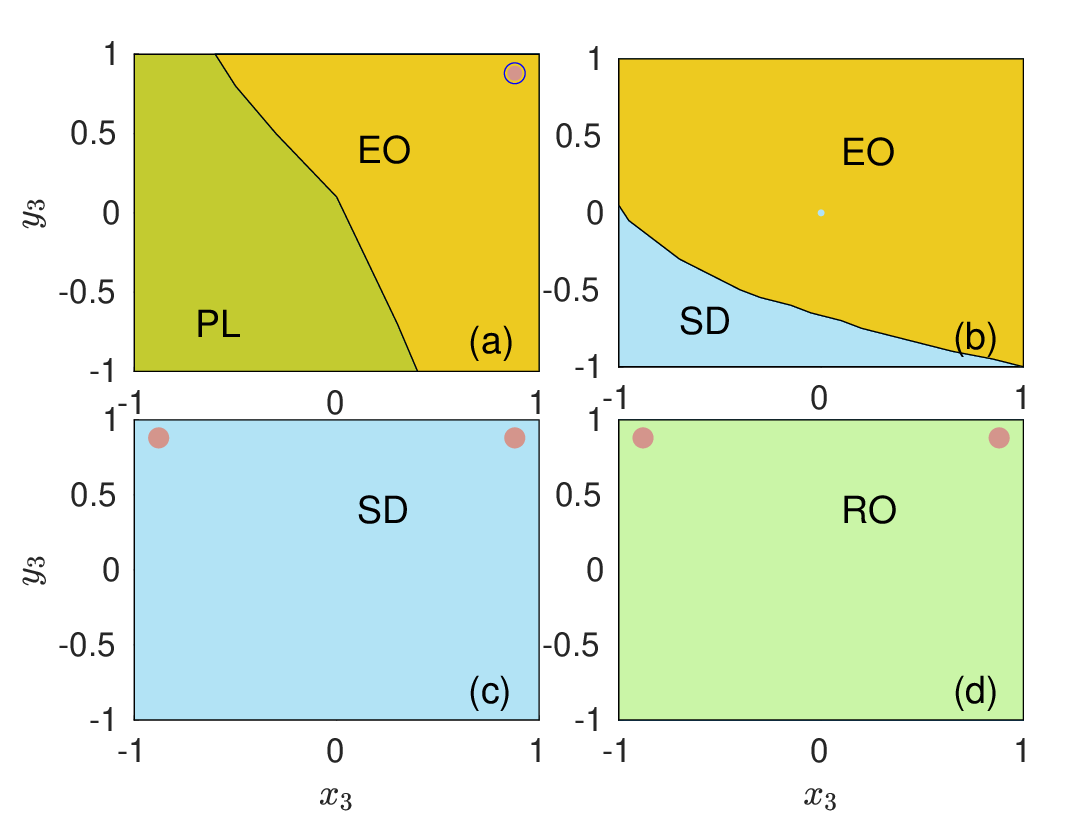}
    	\caption{Basin of attraction for a system of  globally coupled identical SL oscillators with $N=3$ and $\omega=4.0$ governed by Eq.~\ref{higher_order_sl}. (a) synchronized region $\varepsilon=1.0$, (b) Hysteresis region at $\varepsilon=2.7$, (c) Solitary death region at $\varepsilon=4.0$, (d) Revival of oscillation at $\varepsilon=7.0$. }
    \label{fig:basin}
\end{figure}

 \textit{Phase locked (PL) state:} Starting from a set of random initial phases in the forward direction  (increasing $\varepsilon$), we initially encounter a state where the oscillators are phase locked. All the nodes evolve on the same limit cycle but with different phases (Fig.~\ref{fig:figure1}(b)) and are elliptical in the coordinate space. The existence of this state depends on the properties  of the initial conditions. If the angular width of the distribution ($\delta$) is small, this state will not be observed (Fig.~\ref{fig:figure2}(a)). In the bifurcation diagram (Fig.~\ref{fig:basin}), this state is represented by the stable limit cycle with the amplitude of $x_1$ around 2.
 
\begin{figure}[t]
\includegraphics[height=6.0cm,width=\linewidth]{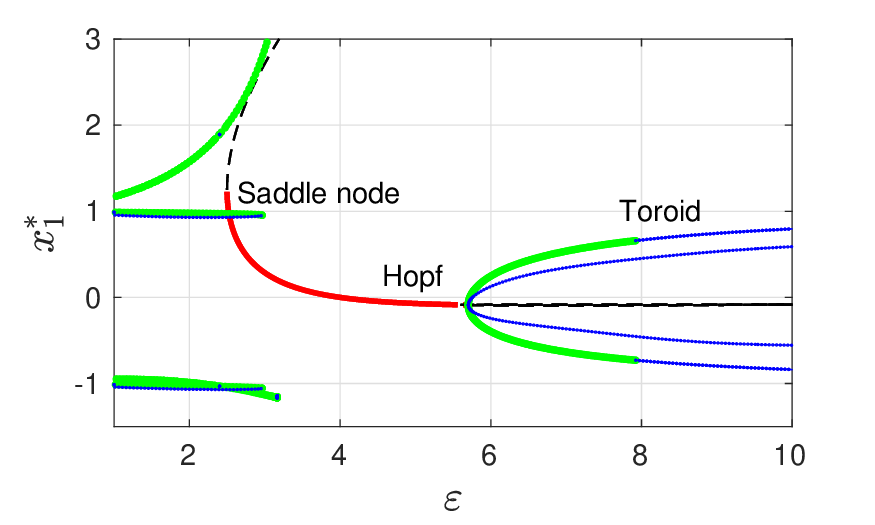}
    \caption{Bifurcation diagram of coupled identical Stuart-Landau oscillator plotted using XPPAUT \cite{Xppaut} for $\omega=4.0$ and $N=3$. The stable oscillatory state is depicted by a green circle, while the unstable oscillatory state is  depicted by a blue circle. A stable steady state is represented by a red solid line while an unstable steady state is represented by the black dashed line. 
    }
    \label{fig:bifurcation}
\end{figure}

 \textit{Enhancement of oscillations (EO):} With a further increase in $\varepsilon$,  an abrupt transition to the synchronized state from the phase-locked state is observed which is accompanied by a sudden increase in the amplitude of the oscillations and the trajectory of the oscillators remains no more elliptical (Fig.~\ref{fig:figure1}(c)). The critical coupling for this abrupt transition to synchronization depends on the system's initial state. 
 The bifurcation diagram best explains this shifting phenomenon of the critical coupling at which the transition to a synchronized state occurs. Fig.~\ref{fig:bifurcation} illustrates that two stable limit cycles exist for $\varepsilon<3.1$. The limit cycle representing the PL state has a constant amplitude for different values of $\varepsilon$ with the elliptical shape. The amplitude of oscillators increases with $\varepsilon$ for the other limit cycle, corresponding to the synchronized state. Depending upon the choice of the initial conditions, the system can settle on any of these two limit cycles.
 Note that in the backward direction, the system  always remains in the synchronized state unless it experiences any perturbation.

 \textit{Solitary death (SD) state:} 
 Upon a further increase in $\varepsilon$, the system undergoes a first-order transition to the SD state (explosive death). Only one unstable fixed point exists before the critical $\varepsilon$ ($\varepsilon_{cb}$). At $\varepsilon_{cb}$, due to the higher-order couplings in the system, a new pair of fixed points is born through the saddle-node (limit point) bifurcation, yielding one stable and one unstable branch. The stable branch corresponds to the solitary death state and it loses stability when $\varepsilon$ increases beyond a certain value. Before that until $\varepsilon_{cf}$, this stable fixed point coexists with two other stable limit cycles. This regime is depicted as the hysteresis loop whose width increases with an increase in the value of $\omega$.
    The numerical simulations indicate that all the oscillators settle to a common fixed point away from the origin. The position of the fixed points depends on $w$ and $k$ and is given by $
        x^{*1}=-\frac{-\omega-2\varepsilon y^{*}+\sqrt{(\omega+2\varepsilon y^*)^2+4y^{*}(y^*-y^{*3})}}{2y^{*}}, $ $ y^{*1}=-\frac{\omega}{\varepsilon}$
and, $ x^{*2}=\frac{\omega+2\varepsilon y^{*}+\sqrt{(\omega+2\varepsilon y^{*})^2+4y^{*}(y^{*}-y^{*3})}}{2y^{*}},$ $ y^{*2}=-\frac{\omega}{\varepsilon}$ along with the preexisting fixed point $x^{*3}=0$, $y^{*3}=0$.
    Next, the characteristic equation for the Jacobian can be written in the form of,
    \begin{align}\label{jaco_general}
    |I\lambda- M| = \left|
\begin{array}{ccccc}
   \ M_1+F_1 & . & . & F_1 \\
   \ F_2 & M_2+F_2 & . & . \\
   \ F_i & . & M_i+F_i & . \\
   \ F_N & . & . & M_N+F_N \\
\end{array}
\right|
\nonumber
\end{align}
where,
$M = 
\begin{pmatrix}
    \lambda-1+3x^2+y^2 & +\omega-2xy\\
    +\omega-2xy & \lambda-1+x^2+3y^2
\end{pmatrix}
$
and,
$F  = \frac{2\varepsilon}{N}
\begin{pmatrix}
    x & y\\
    y & x
\end{pmatrix}
$.
The characteristic equation of these types of solutions is given by \cite{sanket},
\begin{equation}\label{master} \Pi_{i=1}^N|M|=0\;\;\text{and}\;\;|I_2+\sum_{i=1}^{N}\frac{adj(M)F}{|M|} |=0
\end{equation}
The fixed point $x^{*1}$ is unstable for all the values of $\varepsilon$ and $\omega$, confirming the simulation results. We focus on the following eigenvalues for $x^{*2},y^{*2}$ to get the stability condition for the SD state. 
\begin{equation}\label{eqn:eig1}
    \lambda_{1,2}=1-\frac{2\omega^2}{\varepsilon^2}-\frac{\varepsilon^2\eta^2}{2\omega^2}\pm\sqrt{-\omega+\frac{\omega^4}{\varepsilon^4}+\frac{\eta^2}{2}+\frac{\varepsilon^4 \eta^4}{16\omega^4}}
\end{equation}
\begin{equation}\label{eqn:eig2}
    \lambda_{3,4}=1-\frac{2\omega^2}{\varepsilon^2}-\frac{\varepsilon^2\eta}{\omega}-\frac{\varepsilon^2\eta^2}{2\omega^2}\pm\sqrt{-\omega+\frac{\omega^4}{\varepsilon^4}+\frac{\eta^2}{2}+\frac{\varepsilon^4 \eta^4}{16\omega^4}}
\end{equation}
where,
$
\eta=-\omega+\sqrt{\omega^2-\frac{4\omega}{\varepsilon}(-\frac{\omega}{\varepsilon}+\frac{\omega^3}{\varepsilon^3})}
$
The real part of these eigenvalues (Eq.~\ref{eqn:eig1}) must be negative for the fixed point to be stable, which provides us with the conditions $\varepsilon<\sqrt{\frac{1+4\omega^2}{2}}$,  the upper bound for the stability of the fixed point. Similarly, the lower bound is derived by using the fact that the real part of Eq.~\ref{eqn:eig2} is less than zero and consequently $\varepsilon > \sqrt{-2 + 2\sqrt{1+\omega^2}}$. According to these stability conditions, when $\omega=4.0$, we get $2.5<\varepsilon<5.7$, which are in complete agreement with the numerical results (Fig.~\ref{fig:figure1}). Further,  the phase diagram in the parameter space
$(\varepsilon, \omega)$ Fig.~\ref{fig:phase_space} depicts that the analytical conditions match those calculated numerically.

 \textit{Revival of Oscillation (RO):} In the forward direction, once a death state is reached, it persists in an increase in $\varepsilon$. In the backward direction, starting from a set of random initial conditions, an oscillatory state is achieved with the decrease in $\varepsilon$. The fixed point corresponding to the SD state does lose its stability at critical $\varepsilon$; however, in the forward direction, we change $\varepsilon$ adiabatically, the oscillators stay at the fixed point, and the unstable  fixed point keeps getting manifested. Whereas, if we do not set the initial condition  corresponding to a fixed point solution  (as in the case of backward direction), an oscillatory state is achieved at critical $\varepsilon$. This state is, however, not simply elliptic in nature; rather resembles more like a torus. The bifurcation diagram points out that the stable fixed point loses its stability via Hopf bifurcation yielding an unstable fixed point and a stable limit cycle. This stable limit cycle again loses its stability via toroid bifurcation to become torus \cite{SM}. This torus rotates around an unstable limit cycle illustrated in the bifurcation diagram \ref{fig:bifurcation}.

\paragraph{\bf{Sensitivity to initial conditions}} 
Sensitivity to the initial conditions in the model (Eq.~\ref{eqn:SL_high}) is first reflected in the critical point of the transition getting affected by $\delta$.
When we confine the initial condition of the system to a small part of the circle, the system is fully synchronized at a very small $\varepsilon$ (Fig.~\ref{fig:figure2}(a)). However, unlike the earlier case, as we decrease the value of $\delta$ (Fig.~\ref{fig:figure2}(b-d)), the critical point shifts towards the right. Consequently, for $\delta=1.0$, there exists no forward synchronization, and as a result, there exists no hint of the SD state in this system. Similarly, in the revival of the oscillation state, if we start simulations close to the unstable fixed point, the system remains in the SD state; else, it goes to the oscillatory state. Furthermore, it can be understood that the hysteresis region is a bi-stable one in which depending on the initial conditions, the system goes to the synchronized or OD state. Fig.~\ref{fig:figure2} illustrates the dependence of the system's steady state on the initial conditions. Since both the probable states in this region satisfy the condition that $x_i=x_j$ and $y_i=y_j$ $\forall i,j$ we have assumed $x_i=x_j=x_3$ and $y_i=y_j=y_3$.



\paragraph{\textbf{Impact of change in the $\omega$  value:}} Fig.~\ref{fig:phase_space} depicts the behaviour of the order parameters with changes in $\varepsilon$ for different $\omega$ values. Upon increasing $\omega$, while both the forward and backward critical coupling strengths corresponding to SD shift towards the right, $\varepsilon_{cf}$ shifts much larger than $\varepsilon_{cb}$, and consequently, the width of the hysteresis increases. Additionally, the stability region for SD state also increases with an increase in intrinsic frequency $\omega$.
Note that nonidentical oscillators in this setup do not get any interesting emerging dynamics.

\begin{figure}
\includegraphics[height=4.0cm,width=0.45\textwidth]{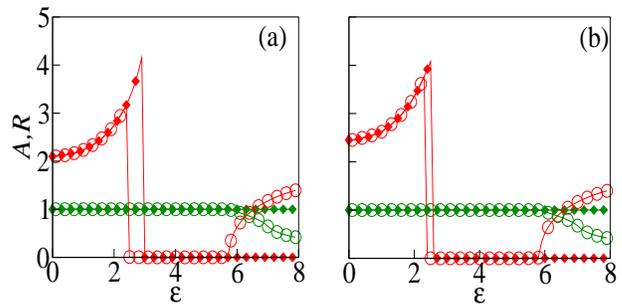}
    \caption{$A,R$ vs $\varepsilon$ for a system of globally coupled identical SL oscillators  for different values of $\varepsilon_p$ (a) $\varepsilon_p=0.1$ (b) $\varepsilon_p=0.6$. $\delta=1.1, N=1000$;  Red diamond - $A$ in the forward direction, green circle - $A$ in the backward direction, green diamond -$R$ in the forward direction, blue circle- $R$ in the backward direction. }
    \label{fig:pairwise_high}
\end{figure}

\paragraph{\textbf{Introduction of pairwise couplings:}} Next, we add pair-wise coupling along with the triadic coupling  in the following manner; 
\begin{equation}\label{eqn:pair+higher_sl}
	\dot z_j(t)=(1-|z_j(t)|^2)z_j+i\omega z_j + \frac{\varepsilon_p}{N}\sum_{k=1}^{N} z_k  + \frac{\varepsilon}{N^2}\sum_{k=1}^{N} \sum_{l=1}^{N} z_k z_l
\end{equation}
Where $\varepsilon_p$ is the pairwise coupling strength.
Fig.~\ref{fig:pairwise_high} indicates that even for small values of $\varepsilon_p$, the PL state vanishes and synchronization is achieved. Moreover, with the introduction of pair-wise coupling, the hysteresis width decreases with an increase in $\varepsilon_p$.

\paragraph{\textbf{Conclusion:}}
 This Letter investigates globally coupled identical oscillators with higher-order interactions. We propose a scheme for incorporating higher-order interactions, which can not be decomposed into lower-order interactions, and also provides a physical meaning to the oscillator systems in their polar coordinate version.   
 We report the emergence of a coupling-dependent SD state, a single stable quenched state arising from the higher-order coupling. This state might be relevant for real-world complex systems, where a single stabilization point is wanted, and can be set using the coupling strength.  
 Moreover, such a coupling scheme yields both explosive death and an abrupt jump to synchronization along with a surge in the amplitude and revival of oscillation in the form of a torus. The critical coupling strength for the transition to synchronization depends on the initial condition's angular spread. Although this property does not seem to be a repercussion of higher-order coupling, other phenomena of the surge in amplitude with synchronization, hysteresis and revival of oscillations seem to result from the coupling scheme. 
 Further, the surge in the amplitude just after the synchronization resembles the pre-ictal regime in which synchronization is accompanied by  an increase in brain activity, 
 which is further followed by PGES (Post-ictal generalized epileptic seizure) corresponding to a considerable suppression of brain activity \cite{seizure_sync, postic_2, kura_brain}. These states can be compared to the EO and SD states manifested by Eq.~\ref{eqn:SL_high}. Moreover, at the end of PGES, the brain might return to a normal state \cite{revival_eg} which resembles the RO state discussed here. Further, this Letter only considers triadic  interactions; a straightforward extension is to incorporate other higher-order interactions, such as quadratic and other coupling forms. 
 
\begin{acknowledgments}
    SJ gratefully acknowledges SERB Power grant SPF/2021/000136. 
    We are thankful for the computational facility from the Department of Science and Technology (DST), Government of India, under the FIST scheme (Grant No. SR/FST/PSI-225/2016). We thank Sandipan Pati for Seizure related discussions.
\end{acknowledgments}

\end{document}